\documentclass[prb,a4paper,showpacs,twocolumn,superscriptaddress,longbibliography]{revtex4-2}

\usepackage{textcomp}
\usepackage{amssymb}
\usepackage{amsmath}
\usepackage{amsfonts}
\usepackage{graphicx}
\usepackage{bm}
\usepackage{xcolor}
\usepackage{multirow}
\usepackage{natbib}
\usepackage{hyperref}
\usepackage{mathrsfs}

\begin{document}

\title{Weyl points in the multi-terminal Hybrid Superconductor-Semiconductor Nanowire devices}

\author{E. V. Repin}

\affiliation{Kavli Institute of Nanoscience, Delft University of Technology, 2628 CJ Delft, The Netherlands}

\author{Y. V. Nazarov}

\affiliation{Kavli Institute of Nanoscience, Delft University of Technology, 2628 CJ Delft, The Netherlands}

\begin{abstract}
The technology of superconductor-semiconductor nanowire devices has matured in the last years in the quest for topological quantum computing. This makes it feasible to make more complex and sophisticated devices. We investigate multi-terminal superconductor-semiconductor wires to access feasibility of another topological phenomenon: Weyl singularities in their spectrum. We have found an abundance of Weyl singularities for devices with intermediate size of the electrodes. We describe their properties and the ways the singularities emerge and disappear upon variation of the setup parameters.  

\end{abstract}

\maketitle

\section*{}
Topological properties of solids have been a subject of intense research for last years\cite{Qi,Bernevig}. The prominent examples of topological materials include topological superconductors\cite{Sato_2017} that may host Majorana modes\cite{Alicea_2012}, and Weyl semimetals\cite{ReviewSemimetals} with Weyl points\cite{Weyl1929} in the electron spectrum. Despite a big interest, the fabrication, purification and experimental analysis of topological materials is difficult and challenging\cite{Liu2019}. This motivates a large effort to realize topologically non-trivial quantum states with topologically trivial materials.\cite{Alicea_2012,Liu2019}

\begin{figure}
	\centerline{\includegraphics[width=0.48\textwidth]{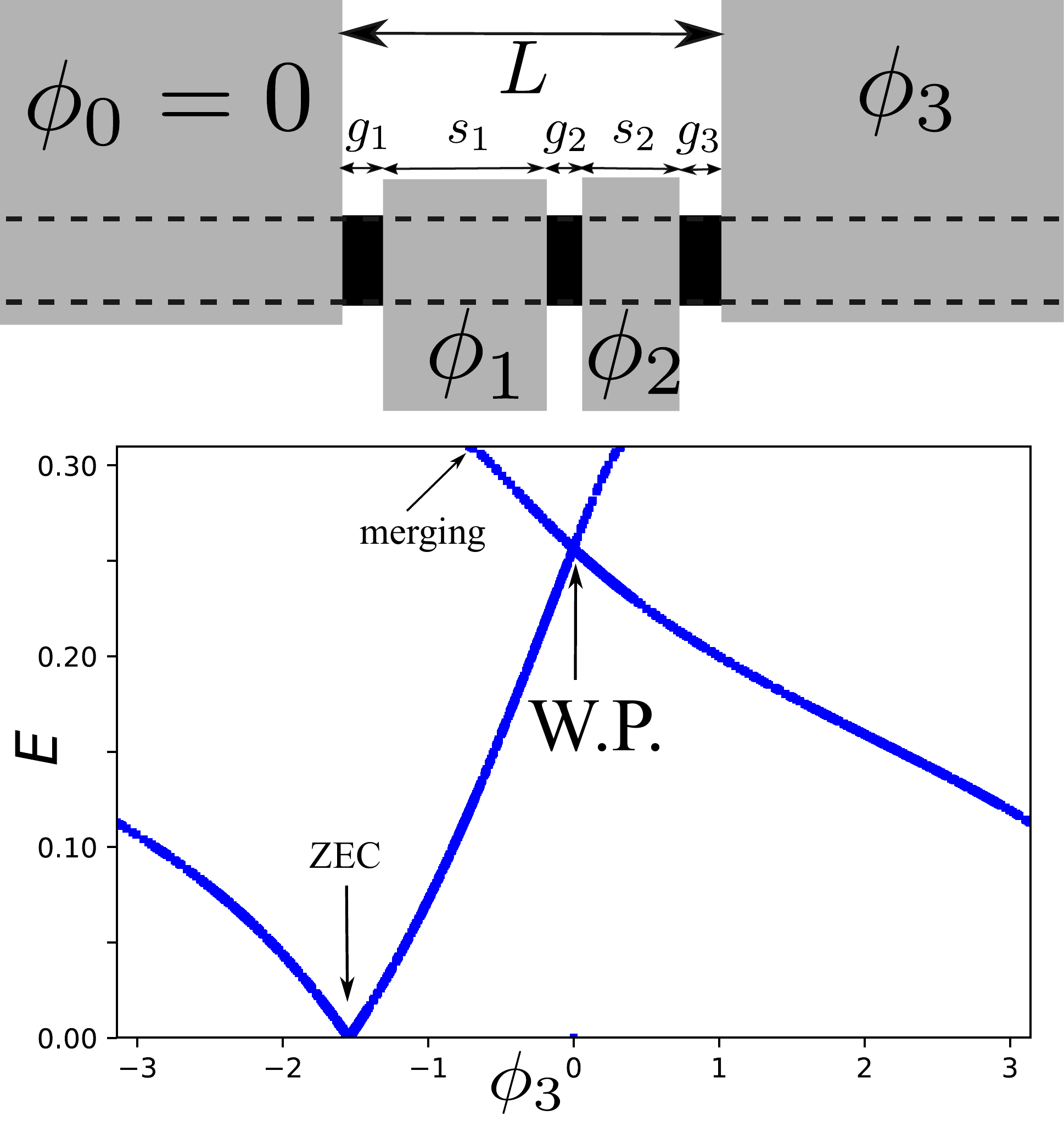}
	}
	\caption{Top: The family of the hybrid superconductor-semiconductor setups under consideration. A long semiconducting nanowire is covered by 4 superconducting leads kept at three independent superconducting phases $\phi_{1,2,3}$. A setup is characterized by overall length $L$ and the lengths of electrodes and gaps, $s_{1,2}, g_{1,2,3}$. Bottom: A typical spectrum of Andreev bound states along the line passing a Weyl point where the bands cross. Other features worth attention are zero-energy crossing that occur at a 2D surface in 3D space of the phases, and merging of the second energy band with the gap edge (top egde of the plot).}
	\label{setup}
\end{figure}

The most known and successful effort of this kind is the realization of zero-energy Majorana states, that can be useful in topological quantum computing \cite{ReviewTopComp2008}, in semiconductor nanowires covered by superconducting electrodes, so-called hybrid superconductor-semiconductor nanowire devices. The first experimental observation \cite{Mourik2012} came only in two years after the first theoretical proposal \cite{Lut}, yet a considerable enhancement of technology was needed for further progress. With the achievement of ballistic superconductivity \cite{Ballistic2017} and experimental verification of topological signatures in Josephson effect \cite{Josephson2019}, the very active sub-field and the technologies in use are mature for next level of experimental sophistication \cite{WimmerReview2020, FrolovReview2020}. One of the interesting directions is the fabrication of the multi-terminal nanowire-based devices.\cite{WimmerReview2020} Recently proposed Andreev molecules \cite{AndreevMolecule2019} that exhibit non-trivial features in the spectrum of the Andreev states \cite{KornichResearch2019,KornichPRB2020} require three superconducting terminals, and the fabrication efforts are underway. In the same manner, one can realize the devices with more terminals.

It has been suggested that the topologically protected spectral singularities - Weyl points - may be realized in multi-terminal superconducting nanostructures\cite{ncomms11167}, potentially, in any nanostructures. The tuning of 3 parameters is required to achieve the singularity, so the minimum number of terminals is four corresponding to three independent superconducting phases.  
The singularity is pinned to zero energy (counted from Fermi level) in the absence of spin-orbit interaction, and is at finite energy distance if spin-orbit interaction is significant \cite{tomohiro}. The topological charge is manifested by transconductance quantization \cite{ncomms11167,Eriksson2017} and can be detected by a spectroscopic measurement \cite{Klees2020}, with some complications brought by the continuous spectrum above the superconducting gap \cite{RepinCont}.
Four-terminal devices have been fabricated in graphene \cite{Draelos2019} and 2D semiconducting structures \cite{pankratova2018multiterminal}. However, the experimental confirmation of Weyl points is not yet available. The presence or absence of Weyl points in any concrete nanostructure depends on the details of scattering that may be difficult to identify and control, and only  6 \% of random scattering matrices provide those. To facilitate the experimental observation and possible applications, it would be good to propose a system where the Weyl points are relatively abundant.

In this Article, we investigate the presence of Weyl points in a spectrum of a single-nanowire four-terminal hybrid semiconducting device of a straightforward design and indeed find many of those. This setup is distinct from that of several nanowires with coupled zero-energy Majorana modes \cite{sakurai2020,stenger2019,
Houzet2019}. In fact, we look for Weyl points at finite energy, where they are present irrespective of the Majorana modes, and find them both in topologically trivial and non-trivial wires.
%spectrum

A typical spectrum with a Weyl point is presented in Fig. \ref{setup}. We set $\phi_{1,2}$ in such a way that the line passes the Weyl point. Other feature of the spectrum is zero-energy crossing (ZEC) \cite{Akhmerov, tomohiro} that occurs at a 2D surface in the 3D space of phases. If the wire is in non-topological regime, there is an even number of ZEC separating the regions with different parity of the ground state. If, as in Fig. \ref{setup}, the wire is in topological regime, the number of crossings may be odd \cite{Lut} manifesting so-called $4\pi$ periodicity. The parity determination requires consideration of the zero-energy state at far ends of the wire \cite{Pikulin}.

%Setup
 We concentrate on a family of setups where a (formally infinite)semiconducting nanowire is covered by 4 separate superconducting films (see Fig.\ref{setup}). The flims are the superconducting leads kept at the corresponding superconducting phases $\phi_0=0,\phi_{1,2,3}$. The widths of 2 intermediate leads $s_{1,2}$ and the gaps between the leads $g_{1,2,3}$ sum up to $L$. A setup of the family is thus characterized by $L$ and five numbers $\vec{s} \equiv [g_1/L,s_1/L,g_2/L,s_2/L,g_3/L]$ summing to 1. We investigate the possibility to realize Weyl points in the 3-dimensional phase space of 3 superconducting phases varying $L$. 
 
 %big and small L
 The wave function of a  Andreev bound state is localized at a typical scale $\xi$. At $L\ll \xi$ we expect no Weyl points since in this case the localized state hardly feels the middle leads and its energy depends on a single parameter only, $E(\phi_3)$. Neither we expect the Weyl points in the opposite limit $L\gg \xi$: in this case, the states are localized in the corresponding gaps $g_i$ with the energies depending on the local phase differences $\phi_i-\phi_{i-1}$, again depending on a single parameter each. Therefore, we expect Weyl points to appear for each setup at $L\sim \xi$. Indeed, for most choices of $\vec{s}$  we find one or more intervals of $L$ where the Weyl points are present, both in topological and non-topological regime. 

%Hamiltonian and how too
We employ the Lutchin-Sau-Das-Sarma Hamiltonian\cite{Lut}; 
\begin{equation}
H=\left(\frac{p^2}{2}-p\sigma_z-\mu\right)\tau_z+{\rm Re}\Delta(x)\tau_x+{\rm Im}\Delta(x)\tau_y+B\sigma_x
\label{ham}
\end{equation}
that we made dimensionless measuring lengths and energies  in units of spin-orbit length and spin-orbit energy, $\tau_i,\sigma_i$ being Pauli matrices in Nambu and spin space, respectively. Here, $\Delta(x)$ is the superconducting order parameter induced in the wire.  We assume a piecewise-constant spacial dependence where $\Delta(x)=|\Delta|e^{i\phi_i}$ under the leads, $\phi_i$ being the phase of the corresponding lead and $\Delta(x)=0$ within the gaps(see Fig.\ref{setup}). The wire is in the topological regime\cite{Lut} provided $|B|>\sqrt{|\Delta|^2+\mu^2}$, otherwise it is in non-topological one.

%Symmetries
The Hamiltonian (\ref{ham}) possesses the usual BdG symmetry $H^{*} = - \sigma_y \tau_y H \tau_y \sigma_y$ that guaranties the symmetry of the spectrum and Weyl points with respect to $E \to -E$. We concentrate at positive energies. Although the Hamiltonian  \ref{ham} is not invariant with respect to time reversal, there is a look-alike extra symmetry
\begin{equation}
H^*(\vec{\phi})=\sigma_x H(-\vec{\phi})\sigma_x
\end{equation}
 relating the Hamiltonians at opposite points $\vec{\phi}$ and $-\vec{\phi}$ in phase space. Therefore, the Weyl points come in pairs of the same charge at opposite points, as for a time-reversible scattering matrix \cite{ncomms11167}. It has been suggested in \cite{ncomms11167} that Weyl points emerge in groups of four conform to conservation topological charge. Here we find notable exceptions from this rule: the Weyl points emerging from the continuous spectrum at the gap edge. 
% We note, however, that there is an interesting possibility of the Weyl points appearing as a pair at positive and negative energy when the Weyl point comes from the continuous spectrum (see Figs.\ref{fignl},\ref{figEsfases},\ref{fig12}). We stress that we consider Weyl points at $E>0$ since the dimensionality of the subspace where $E(\vec{\phi})=0$ is generally not point-like in the model.

\begin{figure}
\centerline{\includegraphics[width=0.48\textwidth]{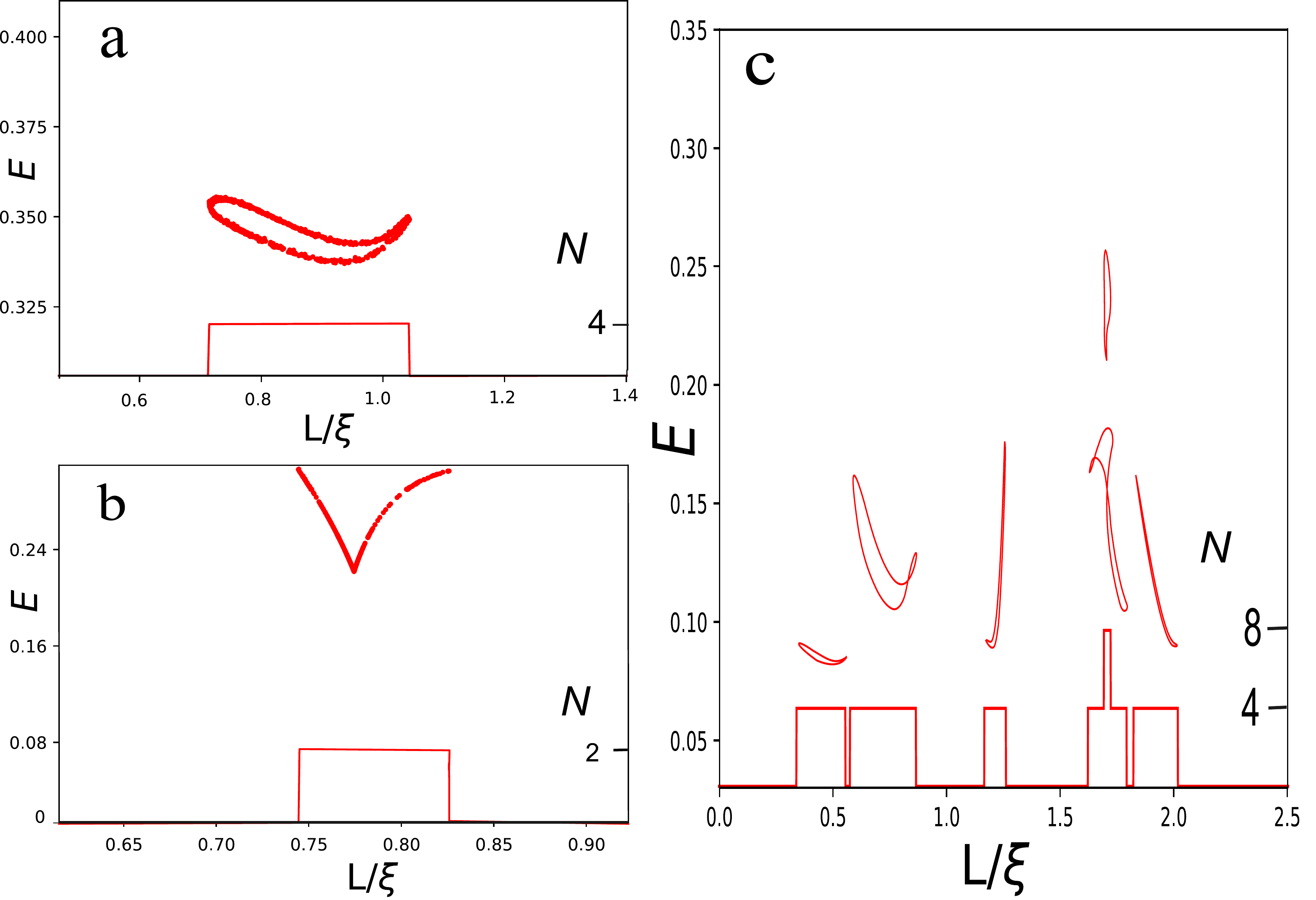} %{fignl.pdf}
}
\caption{
Number and energy dependence of Weyl points for several setups. Non-topological regime: a) $\mu,B,|\Delta|=(1,1,2)$, $\xi\approx 1.07$, $\vec{s}=(0,0.7,0,0.3,0.0)$;
c) $\mu,B,|\Delta|=(1,1,0.9)$, $\xi\approx 2.30$, $\vec{s}=(0.04,0.61,0.09,0.26,0)$
Topological regime: c) $\mu,B,|\Delta|= (0.464,1.144,0.693)$, $\xi\approx 3.26$, $\vec{s}=(0,0.7,0,0.3,0.0)$; 
d) $\mu,B,|\Delta|=(0.951,2.407,1.665)$, $\xi\approx 2.1$, $\vec{s}=(0.05,0.66,0.01,0.28,0.0)$. For b) and c), the gap edge is at the top edge of a plot. For a), the gap edge is at 1.24.}
\label{fig2}
\end{figure}

\begin{figure}
\centerline{\includegraphics[width=0.48\textwidth]{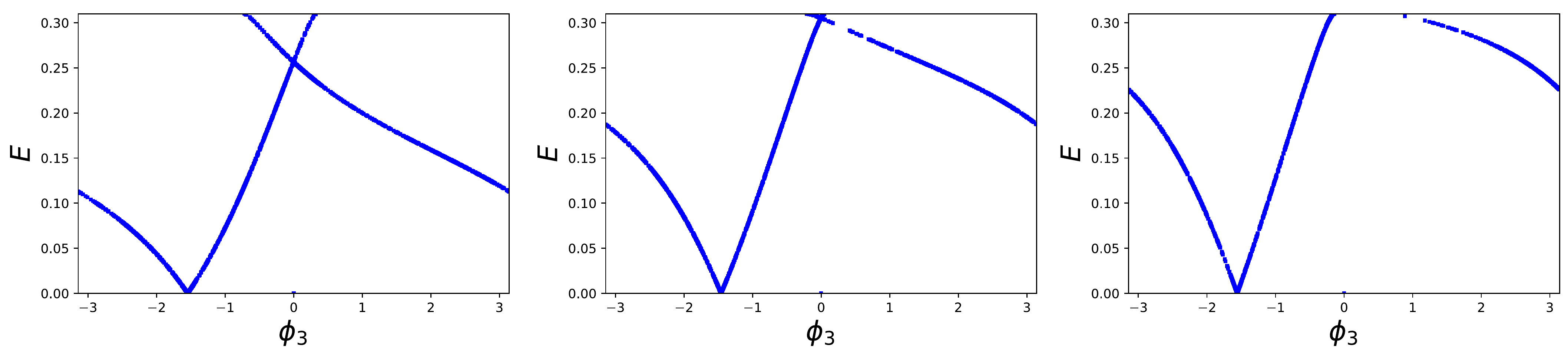} %{fig321.pdf}
}
\caption{Merging of a Weyl point with the gap edge. Choice of the parameters and setup is the same as in Fig. 2c. We plot the spectrum versus $\phi_3$ at a line hitting the Weyl point for three values of $L$. Left: $L/\xi=0.781$, $\phi_1=2.639$, $\phi_2=2.629$ and  , the Weyl point is $E=0.256$ and $\phi_3=0.002$ . Middle: $L/\xi=0.824$, $\phi_1=2.260$, $\phi_2=2.256$, the Weyl point is precisely at the gap edge $E_g =0.31$. Right: $L/\xi=0.829$, there is no Weyl point, no bound state is found in an interval of $\phi_3$.}
\label{fig12}
\end{figure}

\begin{figure}
	\centerline{\includegraphics[width=0.48\textwidth]{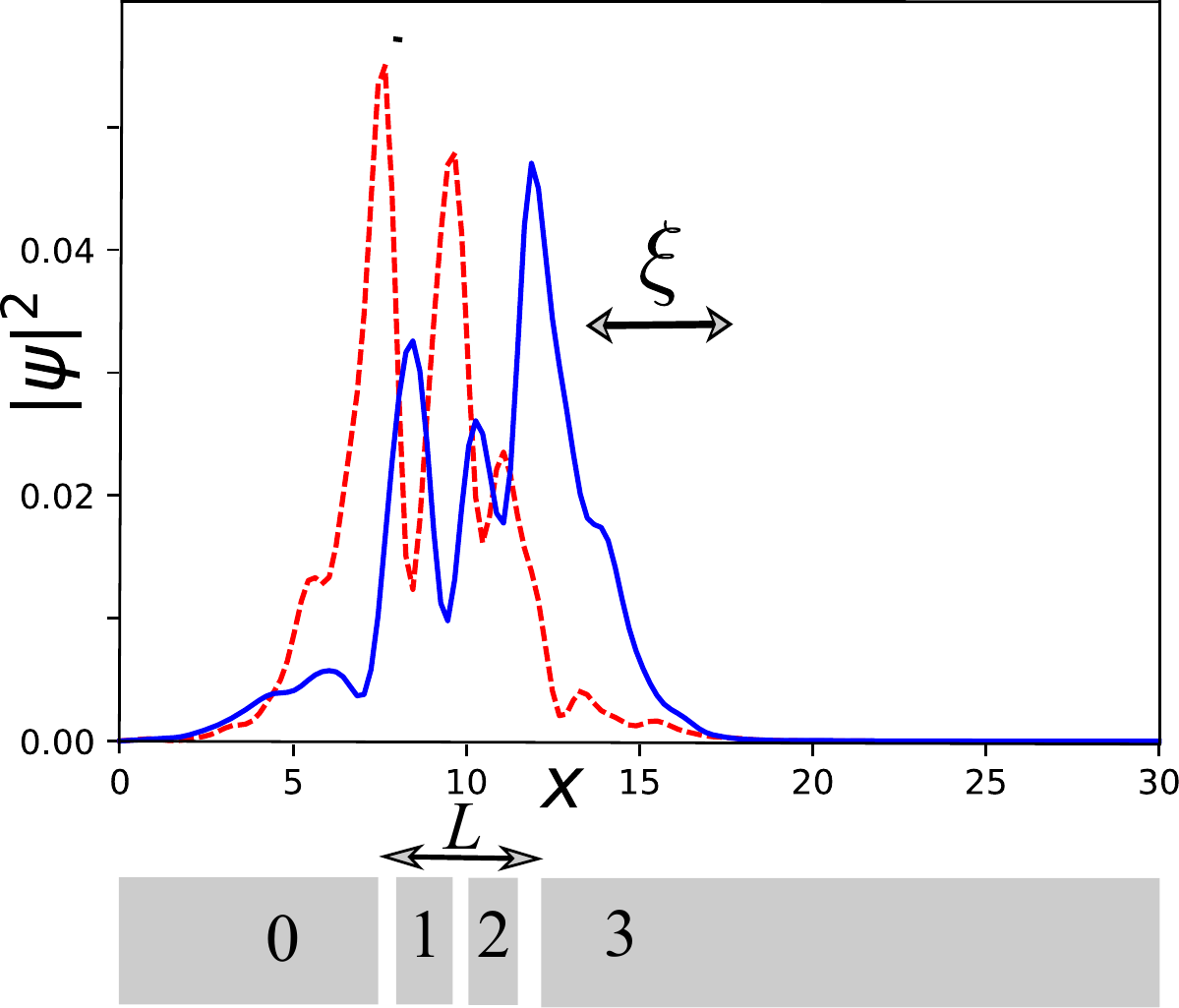} %{fignl.pdf}
	}
	\caption{The densities $\sum_i |\psi(x)|^2_i$ of two degenerate wave functions (solid and dashed curves) at a Weyl point. The wave functions in the degenerate subspace  are chosen to be eigenvectors of the coordinate operator $x$.  The calculations are made for a finite wire of total length  $l=30$, the overlaps with the leads $0-3$ are shown below the plot. The parameters are $B,\mu,|\Delta|=(1,1,0.9)$, corresponding to $\xi =4.0$, $\vec{s}=(1/9,1/3,1/9,1/3,1/9)$. For $L=4.5$, the point is found at $\phi_1,\phi_2,\phi_3=(3.059,-0.448,1.631)$.
	}
	\label{fignl}
\end{figure}

The relevant examples of our numerical results are presented  in Figs.\ref{fig2},\ref{fig12},\ref{fignl}. In  Fig.\ref{fig2} we plot the number and the energies of the Weyl points versus the overall setup length $L$. For each parameter set $\mu,|\Delta|,B$ we normalize $L$ on the localization length $\xi$  that is defined as the slowest decaying exponent under the leads $0,3$.  For all parameters and setups investigated, we find Weyl points in one or several intervals around $L\simeq \xi$. We observe strong dependence of number and energy dependences on the setup details. This is explained by the fact that the Weyl points emerge from complex interference in the setup, the interference pattern being affected by all details.

In non-topological regime (Figs. \ref{fig2}a, \ref{fig2}c) the points come in groups of four. Their energy dependence is seen as a closed curve, a trajectory in $L-E$ space, that does not touch the gap edge. The curves may intersect or self-intersect, the intersection corresponding to the points at the same energy but separated in phase space. The number of Weyl points at given $L$ is 2 times number of intersections of the line $L=const$ with all the curves, as we see in the plots. 
Let us discuss the emergence of Weyl points upon changing $L$ taking Fig. \ref{fig2}a as example. There are no points at $L<0.71$. At $L=0.71$, a pair of close points of opposite topological charges emerges at some phase settings $\vec{\phi}$, with close energies. At the same $L$, another pair emerges near $-\vec{\phi}$, so 4 points appear in total. Upon changing $L$ up to $0.9$, the points got separated in phase settings and energy. As explained in  \cite{ncomms11167, tomohiro}, any 2D plane that separates the points in the phase space, acquires a non-trivial Chern number that is manifested as a quantized transconductance at even parity of the setup. Upon further change of $L$, the points of opposite charges get close together and eventually annihilate at $L= 1.04$. All this is seen as a closed trajectory in $L-E$ space. More complex picture involving multiple trajectories of the same kind (let us call those type A trajectories) is seen in Fig. \ref{fig2}b. 

In topological regime, zero-energy states are formed at the far ends of the wire (this is not detected in our approach that concentrates at the states localized at all electrodes). An example is provided in Fig. \ref{fig2}b. There are no points for $L<0.745$. At $L = 0.745$, a Weyl point emerges from the continuous spectrum at some phase setting $\phi$. The symmetry implies that another point of the same topological charge emerges at $-\phi$, so two points appear in total. Upon changing $L$ the point changes its phase coordinate. It gets lower in energy first, but eventually returns back to the gap edge and disappears at $L=0.781$. Such trajectories begin and end at the gap edge: let us call those type B trajectories.

 We stress that such merging is not compatible with the presence of a continuous band of localized states throughout the Brillouin zone. This is seen from the following topological argument. Let us consider a 2D plane far from the point where the merging occurs. If there is a continuous band throughout the plane, the Chern number is well-defined. However, it must change upon merging. Since the plane is far from the merging point, this is impossible and proves the absence of such band, which also implies the absence of quantized transconductance. Indeed, a detailed view of the spectrum near the Weyl point merging (Fig. \ref{fig12}) shows that the localized states merge with continuous spectrum, and there are regions in the Brillouin zone where no localized state is present.

%Now explain why there is no explanation
In total, we have investigated 12 setups, equal number in topological and non-topological regime. Ten of them have Weyl points in the intervals of $L\simeq \xi$. In several cases, we were not able to trace the whole curve and identify its type. The observation is that the type B trajectory we have seen in the topological regime only. However, no fundamental topological restriction can forbid the type B trajectories in non-topological regime or the type A trajectories in topological regime.  One can see that if one considers a long but finite wire where the overall spectrum is discrete. Such regularization only affects the states at very small energies.  For discrete spectrum, all trajectories are of type A.  Presently, we assume that the observation is valid for the specific family of setups under consideration and is explained by the fact that the boundary conditions near the gap edge in the topological regime are more favourable for merging the localized states with continuum. More detailed research is underway. 

We illustrate the wave functions of the localized states at a Weyl point in Fig. \ref{fignl}. The specifics of the situation is that there are two degenerate wave functions at the point, so eventually one could plot any linear combination of the two. The choice made is as follows: we consider matrix elements of the coordinate operator $x$ in 2-dimensional degenerate subspace, determine and plot the corresponding eigenfunctions. The resulting eigenfunctions are therefore maximally separated in coordinate. We observe the localization of the wave functions at several $\xi$ at the setup, and complex multiple-peak structure that witnesses complex wave interference required for Weyl points. The setup chosen has a mirror symmetry that is however violated by non-symmetric phase settings. Still, the wavefunctions look approximately mirror-symmetric.

To conclude, we have investigated the occurence of Weyl points in the spectrum of Andreev bound states  in a family of realistic device setups where a semiconducting nanowire is covered by four superconducting electrodes. It is feasible to realize such setups experimentally and observe the corresponding topological singularities. For most setups, we find Weyl points for $L \simeq \xi$, that is for setup length of the order of the localization length of the bound states. In experiment, an in situ control of the device length is not feasible. However, it is custom for such devices to utilize a set of gate electrodes to  control $\mu(x)$. We believe that this permits tuning of the device to the region where Weyl points are present.

We observe two types of the Weyl point trajectories. The type A trajectories do not touch the gap edge, and the Weyl points appear in the groups of 4. For type B trajectories, the Weyl points emerge from the gap edge in pairs. We have found the type B trajectory in topological regime only, this should be specific for the family of setups under consideration. 

%%%%%%%%%%%%%%%%%%

\begin{acknowledgements}
We acknowledge useful discussions with Manuel Houzet and Julia Meyer. This research was supported by the European
Research Council (ERC) under the European Union's
Horizon 2020 research and innovation programme (grant
agreement No. 694272).
\end{acknowledgements}

\bibliographystyle{apsrev4-2}
\bibliography{manuscript}

\begin{widetext}

\section*{Supplementary material \label{Sec:App1}}
\subsection{Finding the spectrum}
Here, we outline the method we use to find the spectrum of Andreev bound states in the setup. 
 To start with, we write the  Schrodinger equation for the 4-component eigenfunctions $\psi(x)$ of the Hamiltonian (Eq. \eqref{ham}). This equation is of second-order in $x$-derivative
\begin{equation}
E\psi(x)=[-\frac{\tau_z}{2}\frac{\partial^2}{\partial x^2}+i\sigma_z \tau_z\frac{\partial}{\partial x} +C(x)]\psi(x).
\end{equation}
Here, we define a $4 \times 4$ matrix  $C(x)=B\sigma_x-\mu \tau_z+{\rm Re}\Delta(x) \tau_x+{\rm Im}\Delta(x)$. 
We rewrite it in a form of a first-order differential equation for a new 8-component vector $\Psi(x)$,

\begin{equation}
\frac{\partial \Psi}{\partial x}=\Lambda(x)\Psi,\quad \Psi=\begin{pmatrix}
\psi \\
\partial \psi/\partial x
\end{pmatrix}
\label{LambEq}
\end{equation}
The matrix $\Lambda(x)$ is defined as
\begin{equation}
\Lambda(x)=\begin{pmatrix}
0 & 1\\
2 \tau_z (C(x)-E) & 2i\sigma_z
\end{pmatrix}
\end{equation}
The matrix $\Lambda$ satisfies two symmetries: 
\begin{equation}
\Lambda^*(x)=e^{i\frac{\phi(x)\tau_z}{2}}\sigma_x \Lambda(x) \sigma_x e^{-i\frac{\phi(x)}{2}\tau_z}
\end{equation} 
and 
\begin{equation}
\Lambda \eta_z\sigma_x+ \eta_z\sigma_x\Lambda=0
\end{equation}
where $\eta_z$ is a Pauli matrix in the space $\begin{pmatrix}
\psi \\
\partial \psi/\partial x
\end{pmatrix}$. 
Owing to these symmetries, the eigenvalues of $\Lambda$ come in both complex-conjugated and opposite sign pairs: if $\lambda$ is an eigenvalue, $-\lambda,\lambda^*,-\lambda^*$ are eigenvalues as well. 
There is also an extra symmetry $\Lambda^*(-E) = \eta_z\tau_y \sigma_y \Lambda(E) \sigma_y \tau_y \eta_z$ that guarantees $\lambda(E)=\lambda(-E)$. 
For an infinite lead with a constant $C(x)$,
a pair of purely imaginary eigenvalues at a given energy $E$ indicates a delocalized state at this energy. Since we are interested in the bound states, we restrict our consideration to the energy interval $0<E<E_g$, where the spectral gap $E_g$ is the minimum value of $E$ at which the purely imaginary eigenvalue of $\Lambda$ appears for infinite leads $0, 3$. 

In an interval where $\Lambda$ is constant, the solution of Eq.\eqref{LambEq} reads
\begin{equation}
\Psi(x)=e^{(x-x')\Lambda }\Psi(x')
\label{psisol}
\end{equation}

For the part of the wire covered by the lead 0, where $\Lambda \equiv \Lambda_0$, we need to require the absence of divergent exponents at $x \to -\infty$. For this, let as introduce a projector on the 4-dimensional subspace spanned by the eigenvectors of $\Lambda$. One can compute this projector with using the eigen decomposition of $\Lambda_0$,
\begin{equation}
\Lambda_0 = v_0 \lambda_d v_0^{-1} ;
\end{equation}
where the diagonal $\lambda_d$ is sorted in order of increasing ${\rm Re}\lambda$. With this, 
\begin{equation}
P^-_0 = v_0 P^-_d v_0^{-1},
\end{equation}
where $P^-_d = {\rm diag}\{1,1,1,1,0,0,0,0\}$.

This provides a condition for the wave function $\Psi_0$ at the right end of the interval covered by the lead 0,
\begin{equation}
P^-_0 \Psi_0=0
\label{cond1}
\end{equation}
Similarly, with the projector $P^+_3$ that projects on positive eigenvalues of $\Lambda_3$, we determine the condition on the wave function $\Psi_3$ at the left end of the interval covered by the lead 3,
\begin{equation}
P_3^+\Psi_3=0
\end{equation}
From the other hand, we can implement the relation (\ref{psisol}) throughout the setup to obtain that $\Psi_3 = U \Psi_0$, 
\begin{equation}
U = \exp(g_3 \bar{\Lambda}_3) \exp(s_2 \Lambda_2) \exp(g_2 \bar{\Lambda}_2) \exp(s_1 \Lambda_1) \exp(g_1 \bar{\Lambda}_1);
\end{equation}
$\bar{\Lambda}$ being the $\Lambda$ matrices in the corresponding gaps.
This gives the second condition on the same vector $\Psi_0$:
\begin{equation}
P^+_3 U \Psi_0=0
\label{cond2}
\end{equation}

Next step is to find a proper matrix for which $\Psi_0$ is an eigenvector with zero eigenvalue, $M\Psi_0 =0$. It may seem that any linear combination of Eq. \ref{cond1} and Eq. \ref{cond2} with non-degenerate matrix coefficients would provide such a matrix. However, these linear combinations would also have additional zero eigenvalue eigenvectors that are distinct from $\Psi_0$. To make sure that these additional eigenvectors do not appear, the matrix coefficients in the linear superposition of two conditions should be chosen such that two terms project onto mutually orthogonal subspaces.

The simplest way to achieve this is to multiply Eq. \ref{cond1} with $v^{-1}_0$, Eq. \ref{cond2} with $v^{-1}_3$, and add them up. With this, 
\begin{equation}
M = P^-_d v_0^{-1} + P_d^+ v_3^{-1} U
\end{equation}
and the energy of the bound state is determined from the condition of zero determinant of this matrix. For technical reasons, we prefer to work with an equvalent matrix, $\bar{M} \equiv M v_0$, and solve for
\begin{equation}
{\rm det}(\bar{M})=0
\end{equation}
to find the energies of the bound states. The root finding was implemented as a minimization of the function $F \equiv |{\rm det}(\bar{M})|^2$ over the interval of energies $(0, E_g)$, with subsequent check if zero minimum is achieved. The special properties of the matrix $\Lambda$ guarantees that the zeroes of this complex determinant are achieved at real $E$.

\subsection{Search for Weyl points}
In principle, the presence of a Weyl point can be detected by a thorough scanning of the obtained eigenvalues throughout the whole Brillouin zone of $(\phi_1, \phi_2, \phi_3)$. This, however, is a very time-consuming procedure. 

We automate the search for the Weyl points as follows. It is crucial to note that at a Weyl point the determinat has a double zero, that is, in addition to 
${\rm det}(\bar{M})=0$ the condition $\partial_E {\rm det}(\bar{M})=0$ is also satisfied. So for the search of Weyl points, we fix the setup $\vec{s}$, the parameters $\mu, B, |\Delta|$ and minimize the function
\begin{equation}
F = |{\rm det}(\bar{M})|^2 + |\partial_E {\rm det}(\bar{M})|^2
\end{equation}
in the 5-dimensional parameter space $(E, L, \phi_1, \phi_2, \phi_3)$ starting from a random point and checking if zero minimum is achieved. The output is a set of points in this space that lie at a 1-dimensional manifold. This is how the data plotted in Fig. 2 have been obtained. Once the coordinates of a Weyl point are found, one can compute the spectrum along a line in phase space that passes the point: this is how the plot in Fig. 1 has been obtained. The procedure described also permits finding the wave functions of the localized states. However, for the plots in Fig. 4 we made use of direct diagonalization of a  discrete version of the Hamiltonian \ref{ham}.

\end{widetext}

\end{document}